%
\font\twelverm=cmr10 scaled 1200    \font\twelvei=cmmi10 scaled 1200
\font\twelvesy=cmsy10 scaled 1200   \font\twelveex=cmex10 scaled 1200
\font\twelvebf=cmbx10 scaled 1200   \font\twelvesl=cmsl10 scaled 1200
\font\twelvett=cmtt10 scaled 1200   \font\twelveit=cmti10 scaled 1200
\font\twelvesc=cmcsc10 scaled 1200  
\skewchar\twelvei='177   \skewchar\twelvesy='60
     
     
\def\twelvepoint{\normalbaselineskip=12.4pt plus 0.1pt minus 0.1pt
  \abovedisplayskip 12.4pt plus 3pt minus 9pt
  \belowdisplayskip 12.4pt plus 3pt minus 9pt
  \abovedisplayshortskip 0pt plus 3pt
  \belowdisplayshortskip 7.2pt plus 3pt minus 4pt
  \smallskipamount=3.6pt plus1.2pt minus1.2pt
  \medskipamount=7.2pt plus2.4pt minus2.4pt
  \bigskipamount=14.4pt plus4.8pt minus4.8pt
  \def\rm{\fam0\twelverm}          \def\it{\fam\itfam\twelveit}%
  \def\sl{\fam\slfam\twelvesl}     \def\bf{\fam\bffam\twelvebf}%
  \def\mit{\fam 1}                 \def\cal{\fam 2}%
  \def\sc{\twelvesc}               \def\tt{\twelvett}
  \def\sf{\twelvesf}
  \textfont0=\twelverm   \scriptfont0=\tenrm   \scriptscriptfont0=\sevenrm
  \textfont1=\twelvei    \scriptfont1=\teni    \scriptscriptfont1=\seveni
  \textfont2=\twelvesy   \scriptfont2=\tensy   \scriptscriptfont2=\sevensy
  \textfont3=\twelveex   \scriptfont3=\twelveex  \scriptscriptfont3=\twelveex
  \textfont\itfam=\twelveit
  \textfont\slfam=\twelvesl
  \textfont\bffam=\twelvebf \scriptfont\bffam=\tenbf
  \scriptscriptfont\bffam=\sevenbf
  \normalbaselines\rm}
     

     
\def\beginlinemode{\endmode
  \begingroup\parskip=0pt \obeylines\def\\{\par}\def\endmode{\par\endgroup}}
\def\beginparmode{\endmode
  \begingroup \def\endmode{\par\endgroup}}
\let\endmode=\par
{\obeylines\gdef\
{}}
\def\singlespace{\baselineskip=\normalbaselineskip}

\def\oneandahalfspace{\baselineskip=\normalbaselineskip
  \multiply\baselineskip by 3 \divide\baselineskip by 2}
\def\doublespace{\baselineskip=\normalbaselineskip \multiply\baselineskip by 2}

\newcount\firstpageno
\firstpageno=2
\footline={\ifnum\pageno<\firstpageno{\hfil}\else{\hfil\twelverm\folio\hfil}\fi}
\def\toppageno{\global\footline={\hfil}\global\headline
  ={\ifnum\pageno<\firstpageno{\hfil}\else{\hfil\twelverm\folio\hfil}\fi}}
\let\rawfootnote=\footnote              
\def\footnote#1#2{{\rm\singlespace\parindent=0pt\parskip=0pt
  \rawfootnote{#1}{#2\hfill\vrule height 0pt depth 6pt width 0pt}}}
\def\raggedcenter{\leftskip=4em plus 12em \rightskip=\leftskip
  \parindent=0pt \parfillskip=0pt \spaceskip=.3333em \xspaceskip=.5em
  \pretolerance=9999 \tolerance=9999
  \hyphenpenalty=9999 \exhyphenpenalty=9999 }
\def\dateline{\rightline{\ifcase\month\or
  January\or February\or March\or April\or May\or June\or
  July\or August\or September\or October\or November\or December\fi
  \space\number\year}}
\def\received{\vskip 3pt plus 0.2fill
 \centerline{\sl (Received\space\ifcase\month\or
  January\or February\or March\or April\or May\or June\or
  July\or August\or September\or October\or November\or December\fi
  \qquad, \number\year)}}
     
     
\hsize=6.5truein
\vsize=9.9truein  
\voffset=-1.0truein
\parskip=\medskipamount
\def\\{\cr}
\twelvepoint            
\doublespace            
\overfullrule=0pt       

\def\title                      
  {\null\vskip 3pt plus 0.2fill
   \beginlinemode \doublespace \raggedcenter \bf}
     
\def\author                     
  {\vskip 3pt plus 0.2fill \beginlinemode
   \singlespace \raggedcenter\sc}
     
\def\affil                      
  {\vskip 3pt plus 0.1fill \beginlinemode
   \oneandahalfspace \raggedcenter \sl}
     
\def\abstract                   
  {\vskip 3pt plus 0.3fill \beginparmode
   \singlespace ABSTRACT: }
     
\def\endtopmatter               
  {\endpage                     
   \body}
     
\def\body                       
  {\beginparmode}               
     
\def\head#1{                    
  \goodbreak\vskip 0.5truein    
  {\immediate\write16{#1}
   \raggedcenter \uppercase{#1}\par}
   \nobreak\vskip 0.25truein\nobreak}

\def\beginitems{
\par\medskip\bgroup\def\i##1 {\item{##1}}\def\ii##1 {\itemitem{##1}}
\leftskip=36pt\parskip=0pt}
\def\enditems{\par\egroup}
     
\def\beneathrel#1\under#2{\mathrel{\mathop{#2}\limits_{#1}}}
     
\def\refto#1{$^{#1}$}           
     
\def\references                 
  {\head{References}            
   \beginparmode
   \frenchspacing \parindent=0pt \leftskip=1truecm
   \parskip=8pt plus 3pt \everypar{\hangindent=\parindent}}

\gdef\refis#1{\item{#1.\ }}                     
     
\gdef\journal#1, #2, #3, 1#4#5#6{               
    {\sl #1~}{\bf #2}, #3 (1#4#5#6)}            

\gdef\refa#1, #2, #3, #4, 1#5#6#7.{\noindent#1, #2 {\bf #3}, #4 (1#5#6#7).\rm} 

\gdef\refb#1, #2, #3, #4, 1#5#6#7.{\noindent#1 (1#5#6#7), #2 {\bf #3}, #4.\rm} 

\def\pr{\journal Phys.Rev., }

\def\prl{\journal Phys.Rev.Lett., }
     
\def\jmp{\journal J.Math.Phys., }

\def\endreferences{\body}

\def\endpage                    
  {\vfill\eject}
     
\def\endpaper                   
  {\endmode\vfill\supereject}

\def\ref#1{Ref.~#1}                     
\def\Ref#1{Ref.~#1}                     
\def\[#1]{[\cite{#1}]}
\def\cite#1{{#1}}
\def\(#1){(\call{#1})}
\def\call#1{{#1}}
\def\taghead#1{}
\def\frac#1#2{{#1 \over #2}}
\def\half{{\frac 12}}

\def\12{{1\over2}}

\catcode`@=11
\newcount\r@fcount \r@fcount=0
\newcount\r@fcurr
\immediate\newwrite\reffile
\newif\ifr@ffile\r@ffilefalse
\def\w@rnwrite#1{\ifr@ffile\immediate\write\reffile{#1}\fi\message{#1}}

\def\writer@f#1>>{}
\def\referencefile{
  \r@ffiletrue\immediate\openout\reffile=\jobname.ref%
  \def\writer@f##1>>{\ifr@ffile\immediate\write\reffile%
    {\noexpand\refis{##1} = \csname r@fnum##1\endcsname = %
     \expandafter\expandafter\expandafter\strip@t\expandafter%
     \meaning\csname r@ftext\csname r@fnum##1\endcsname\endcsname}\fi}%
  \def\strip@t##1>>{}}

\def\citeall#1{\xdef#1##1{#1{\noexpand\cite{##1}}}}
\def\cite#1{\each@rg\citer@nge{#1}}	

\def\each@rg#1#2{{\let\thecsname=#1\expandafter\first@rg#2,\end,}}
\def\first@rg#1,{\thecsname{#1}\apply@rg}	
\def\apply@rg#1,{\ifx\end#1\let\next=\relax
\else,\thecsname{#1}\let\next=\apply@rg\fi\next}

\def\citer@nge#1{\citedor@nge#1-\end-}	
\def\citer@ngeat#1\end-{#1}
\def\citedor@nge#1-#2-{\ifx\end#2\r@featspace#1 
  \else\citel@@p{#1}{#2}\citer@ngeat\fi}	
\def\citel@@p#1#2{\ifnum#1>#2{\errmessage{Reference range #1-#2\space is bad.}%
    \errhelp{If you cite a series of references by the notation M-N, then M and
    N must be integers, and N must be greater than or equal to M.}}\else%
 {\count0=#1\count1=#2\advance\count1 by1\relax\expandafter\r@fcite\the\count0,
  \loop\advance\count0 by1\relax
    \ifnum\count0<\count1,\expandafter\r@fcite\the\count0,%
  \repeat}\fi}

\def\r@featspace#1#2 {\r@fcite#1#2,}	
\def\r@fcite#1,{\ifuncit@d{#1}
    \newr@f{#1}%
    \expandafter\gdef\csname r@ftext\number\r@fcount\endcsname%
                     {\message{Reference #1 to be supplied.}%
                      \writer@f#1>>#1 to be supplied.\par}%
 \fi%
 \csname r@fnum#1\endcsname}
\def\ifuncit@d#1{\expandafter\ifx\csname r@fnum#1\endcsname\relax}%
\def\newr@f#1{\global\advance\r@fcount by1%
    \expandafter\xdef\csname r@fnum#1\endcsname{\number\r@fcount}}

\let\r@fis=\refis			
\def\refis#1#2#3\par{\ifuncit@d{#1}
   \newr@f{#1}%
   \w@rnwrite{Reference #1=\number\r@fcount\space is not cited up to now.}\fi%
  \expandafter\gdef\csname r@ftext\csname r@fnum#1\endcsname\endcsname%
  {\writer@f#1>>#2#3\par}}

\def\ignoreuncited{
   \def\refis##1##2##3\par{\ifuncit@d{##1}%
    \else\expandafter\gdef\csname r@ftext\csname r@fnum##1\endcsname\endcsname%
     {\writer@f##1>>##2##3\par}\fi}}

\def\r@ferr{\endreferences\errmessage{I was expecting to see
\noexpand\endreferences before now;  I have inserted it here.}}
\let\r@ferences=\references
\def\references{\r@ferences\def\endmode{\r@ferr\par\endgroup}}

\let\endr@ferences=\endreferences
\def\endreferences{\r@fcurr=0
  {\loop\ifnum\r@fcurr<\r@fcount
    \advance\r@fcurr by 1\relax\expandafter\r@fis\expandafter{\number\r@fcurr}%
    \csname r@ftext\number\r@fcurr\endcsname%
  \repeat}\gdef\r@ferr{}\endr@ferences}


\let\r@fend=\endpaper\gdef\endpaper{\ifr@ffile
\immediate\write16{Cross References written on []\jobname.REF.}\fi\r@fend}

\catcode`@=12

\citeall\refto		
\citeall\ref		%
\citeall\Ref		%

\def\a{{\alpha}}
\def\b{{\beta}}

\def\s{\sigma}
\def\half{{1 \over 2}}
\def\ra{{\rangle}}
\def\la{{\langle}}

\def\ih{{i \over \hbar}}

\def\bp{{\bar p}}
\def\bq{{\bar q}}
\def\bx{{\bar x}}

\def\ria{{\rightarrow}}

\centerline{\bf The Post-Decoherence Density Matrix Propagator}
\centerline{\bf for Quantum Brownian Motion}

\vskip 0.3in
\author Jonathan Halliwell and Andreas Zoupas
\affil
Theory Group, Blackett Laboratory
Imperial College, London SW7 2BZ
UK
\vskip 0.5in
\centerline {\rm Preprint IC 96--95/67, August, 1996}
\vskip 0.2in 
\centerline {\rm Submitted to {\sl Physical Review D}}

\abstract  
{ 
Using the path integral representation of the density matrix
propagator of quantum Brownian motion, we derive its asymptotic form
for times greater than the so-called localization time,
$ (\hbar / \gamma k T )^{\half}$, where
$\gamma$ is the dissipation and $T$ the temperature of the thermal
environment. The localization time is typically greater than the
decoherence time, but much shorter than the relaxation time,
$\gamma^{-1}$. We use this result to show that the reduced density
operator rapidly evolves into a state which is approximately
diagonal in a set of generalized coherent states.  We thus
reproduce, using a completely different method, a result we
previously obtained using the quantum state diffusion picture ({\sl
Phys.Rev.}{\bf D52}, 7294 (1995)). We also go beyond this earlier
result, in that we derive an explicit expression for the weighting
of each phase space localized state in the approximately diagonal
density matrix, as a function of the initial state. For sufficiently
long times it is equal to the Wigner function, and we confirm that the
Wigner function is positive for times greater than the localization time
(multiplied by a number of order $1$).
} 
\endtopmatter

\head{\bf 1. Introduction}

One of the simplest open systems that is amenable to straightforward
analysis is the quantum Brownian motion model. This model consists
of a non-relativistic point particle, possibly in a potential, 
coupled to a bath of harmonic oscillators in a thermal state. The
quantum Brownian motion model has been used very extensively in
studies of decoherence and emergent classicality 
(see for example, Refs.[\cite{QBM,CaL,DoH,Gal,HPZ,PHZ,UnZ,Zur}]).

In the simplest case of a free particle of mass $m$
in a high temperature bath, with negligible dissipation
the master equation for the reduced density matrix $\rho (x,y) $
of the point particle is,
$$
{ \partial \rho \over \partial t}
= { i \hbar  \over 2 m} \left(
{ \partial^2 \rho \over \partial x^2} -
{\partial^2 \rho \over \partial y^2} \right)
-  \half a^2 (x-y)^2 \rho
\eqno(1.1)
$$
where $ a^2 = 4 m \gamma k T / \hbar^2  $. (More general forms of
this equation, together with the derivations of it may be
found in many places. See, for example, Refs.[\cite{CaL,HaT,HPZ}]).

One of the most important properties of (1.1) (and also
its more general forms) is that the density operator tends
to become approximately diagonal in both position and momentum
after a short time. This has been seen in
numerical solutions and in the evolution of particular
types of initial states for which analytic solution
is possible [\cite{JoZ,PHZ,PaZ,UnZ,ZHP,Zur,Zur3b,Zur4,Zur5}].

A more precise demonstration of this statement was given in 
Ref.[\cite{HaZ}]
by appealing to an alternative description of open systems known as
the quantum state diffusion picture [\cite{GP1,GP2,GP3,Dio3,DGHP}]. 
In that picture, the density operator
$\rho$ satisfying (1.1) is regarded 
as a mean over a distribution of pure state
density operators, 
$$
\rho = M | \psi \ra \la \psi | 
\eqno(1.2)
$$
where $M$ denotes the mean (defined below),
with the pure states evolving according to a
non-linear stochastic Langevin-Ito equation,
which for the model of this paper is,
$$
| d \psi \ra = -\ih H |\psi \ra dt 
- \half \left( L - \la L \ra \right)^2 | \psi \ra dt
+ \left( L - \la L \ra \right) | \psi \ra \ d \xi (t)
\eqno(1.3) 
$$
for the normalized state vector $| \psi \ra $, where
$ H = p^2 / 2m $ and $ L = a {\hat x} $.
Here, $d \xi$ is a complex differential random
variable representing a complex Wiener process. The linear and
quadratic means are,
$$ 
M [ d \xi d \xi^* ] = \ dt, \quad
M[ d \xi d \xi ] = 0, \quad M [ d\xi ] = 0
\eqno(1.4)
$$

The appeal of this picture is that the solutions to the
stochastic equation (1.3) appear to describe the expected
behaviour of an individual history of the system, and have
been seen to correspond to single runs of laboratory experiments.
For example, for the quantum Brownian motion model, the solutions
tend to phase space localized states of constant
width whose centres undergo classical Brownian motion 
[\cite{Dio3,HaZ,GaG,SaG,BPSG}].
The timescale of this process, the localization time, is at
slowest of order $ ( \hbar / \gamma k T )^{\half} $, which is
the timescale on which the thermal fluctuations overtake the
quantum fluctuations [\cite{HZ,AndH,AnH}]. For an initial superposition of
localized states a distance $\ell$ apart, localization initally
proceeds on a much shorter timescale, of order 
$ \hbar^2 / ( \ell^2 m \gamma k T ) $ (which
is often called the decoherence time [\cite{Zur,Zur4}]), 
thereafter going over to the slower timescale above.

For us, the interesting feature of the quantum state diffusion
picture is that it gives some useful information about the
form of the density operator on time scales greater than
the localization time.
Given a set of localized phase space solutions $ | \Psi_{pq} \ra $, 
the density operator may be reconstructed via (1.2). 
This, it may be shown [\cite{HaZ}], may be written explicitly as
$$
\rho = \int dp dq \ f(p,q,t) | \Psi_{pq} \ra \la \Psi_{pq} |
\eqno(1.5)
$$
Here, $f(p,q,t)$ is a non-negative, normalized
solution to the Fokker-Planck equation
describing the classical Brownian motion undergone by the
centres of the stationary solutions. 
This is therefore an explicit, albeit indirect,
demonstration of the approach to approximately phase space
diagonal form on short time scales. 

The above demonstration was described by us in detail in Ref.[\cite{HaZ}].
However, we were not able to deduce an explicit form for the 
function $f(p,q,t)$ using the quantum state diffuion picture.
That is, we know that it is a solution to the Fokker-Planck
equation, but it was not clear how to pick out the particular
solution corresponding to a particular initial density operator. 
Intuitively, it is clear that $f(p,q,t)$ is something like
the Wigner function of the initial state, coarse-grained
sufficiently to make it positive, evolved forwards in time,
and with the interference terms thrown away. We would
like to be able to show this explicitly.

The aim of the present paper is to derive the form (1.5)
for times greater than the localization time
directly from the path integral representation of the density matrix
propagator corresponding to (1.1), without using the quantum state
diffusion picture. As we shall see, this derivation has the advantage
that it gives an explicit expression for $f(p,q,t)$. In particular,
we shall show that $f(p,q,t)$ coincides with the Wigner function
$W_t (p,q)$ of the density operator at time $t$, for sufficiently
large times.

\head{\bf 2. The Density Matrix Propagator}

The solution to the master equation (1.1) may be written in terms
of the propagator, $J$,
$$
\rho_t (x,y) = \int dx_0 dy_0 \ J (x,y,t|x_0,y_0,0) \ \rho_0 (x_0,
y_0)
\eqno(2.1)
$$
(see, for example, Refs.[\cite{CaL,AnH}] for further details of
the quantum Brownian motion model).
The propagator may be given in general by a path integral
expression, which for the particular case considered here is
$$
J (x_f, y_f, t | x_0, y_0, 0 ) = \int {\cal D} x {\cal D} y 
\ \exp \left( { i m \over 2 \hbar } \int dt ( {\dot x}^2 - {\dot y}^2 )
- {a^2 \over 2}  \int dt (x-y)^2 \right)
\eqno(2.2)
$$
This is readily evaluated, with the result,
$$
\eqalignno{
J (x_f, y_f, t | x_0, y_0, 0 ) = & 
\exp \left(
{ i m \over 2 \hbar t} \left[ (x_f-x_0)^2 - (y_f-y_0)^2 \right]
\right.
\cr
& \left.
- {a^2 t \over 6}  
\left[ (x_f -y_f)^2 + (x_f - y_f) (x_0 - y_0) + (x_0 - y_0)^2
\right] \right)
&(2.3) \cr }
$$
(For convenience we will ignore prefactors in what follows.
They may be recovered where required by appropriate normalizations.)

The main result of the present paper comes from the simple
observation that the real part of the exponent in the path integral
(2.2) may be written
$$
\exp \left( - {a^2 \over 2} \int dt \ (x-y)^2 \right)
= \int {\cal D} \bx \ \exp \left(
- a^2 \int dt \ ( x - \bx )^2 - a^2 \int dt \ (y - \bx )^2 \right)
\eqno(2.4)
$$
The path integral representation of the propagator may therefore be
written,
$$
J (x_f, y_f, t | x_0, y_0, 0 ) = \int {\cal D} \bx
\ K_{\bx} (x_f,t|x_0,0) \ K_{\bx}^* (y_f,t|y_0,0)
\eqno(2.5)
$$
where
$$
K_{\bx}(x_f,t|x_0,0) = \int {\cal D} x 
\ \exp \left( { i m \over 2 \hbar} \int dt
\ {\dot x}^2 - a^2 \int dt \ ( x - \bx)^2 \right)
\eqno(2.6)
$$
For a pure initial state, $\rho_0 (x,y) = \Psi_0 (x) \Psi_0^* (y) $,
the density operator at time $t$ may therefore be written,
$$
\rho_t (x,y) = \int {\cal D} \bx \ \Psi_{\bx} (x,t) \Psi^*_{\bx}
(y,t)
\eqno(2.7)
$$
where the (unnormalized) wave function $\Psi_{\bx}$ is given by
$$
\Psi_{\bx} (x_f,t) = \int d x_0
\ K_{\bx} (x_f,t|x_0,0) \ \Psi_0 (x_0)
\eqno(2.8)
$$
(Wave functions of this type often appear in discussions
of systems undergoing continuous measurement 
[\cite{CaM,Men,Dio2,GRW}].)

Our strategy is to first evaluate the quantity
$ K_{\bx}$, examine is asymptotic form for times greater than
the localization time, and then
use it to reconstruct the density matrix propagator, $J$.
The reason we expect this to yield the desired result is
that up to normalization factors and ignoring the fact that
$\bx$ is real not complex, Eq.(2.8) 
is essentially the solution to the Langevin--Ito equation, (1.3),
so the phase space localization effect should be visible
in its long time limit. Moreover, Eq.(2.7) is the analogue
of (1.2) or (1.5), so by reorganizing the functional integral
over $\bx (t)$, we might reasonably expect to derive (1.5).

The path integral (2.6) is essentially the same as that for
a harmonic oscillator coupled to an external source, with the
complication that the frequency is complex. The path integral
is therefore readily carried out (see Ref.[\cite{Sch}], for example), 
with the result,
$$
\eqalignno{
K_{\bx}(x_f,t|x_0,0) =  N & \exp \left( \ih c_1 (x_f^2 + x_0^2 )
+ \ih c_2 x_f x_0 
\right.
\cr
& \left.
+ c_3 x_f + c_4 x_0 + c_5 - a^2 \int_0^t ds \ \bx^2 (s) \right)
&(2.9) \cr }
$$
where, 
$$
\eqalignno{
c_1 &= { m \omega \cos \omega t \over 2 \sin \omega t} 
&(2.10)\cr
c_2 &= - { m \omega \over \sin \omega t}
&(2.11)\cr 
c_3 &= {2 a^2 \over \sin \omega t} \int_0^t ds \ \bx (s)
\ \sin \omega s
&(2.12)\cr
c_4 &= {2 a^2 \over \sin \omega t} \int_0^t ds \ \bx (s)
\ \sin \omega (t-s)
&(2.13)\cr
c_5 &= { 4 i \hbar a^2 \over m \omega \sin \omega t}
\int_0^t ds \int_0^s ds' \ \bx (s) \bx (s')
\ \sin \omega (t-s) \ \sin \omega s'
&(2.14) \cr }
$$
Here $\omega = \a (1-i) $ and 
$$
\a = \left({ \hbar a^2 \over 4 m } \right)^{\half}
= \left( { \gamma k T \over \hbar} \right)^{\half}
\eqno(2.15)
$$

The timescale of evolution according to (2.9) is therefore
$\a^{-1}$, which coincides with the localization time discussed in
Ref.[\cite{HaZ}]. The asymptotic properties of $K_{\bx}$ are now easily seen.
As $t \ria \infty$, $ c_2 \ria 0 $ and 
$c_1 \ria \half m \a (1+i) $ like $ e^{- \a t } $.
Since $c_2 \ria 0 $, the propagator $ K_{\bx}$ factors
into a product of functions of $x_0$ and $x_f$.
The wave function (2.8) therefore ``forgets'' its 
initial conditions and becomes proportional to a Gaussian of
the form
$$
\exp \left( \ih c_1 x_f^2 + c_3 x_f  \right)
\eqno(2.16)
$$
on a timescale $\a^{-1}$. This is in complete 
agreement with the quantum state diffusion picture
analysis of Refs.[\cite{Dio3,HaZ}].

Now introduce
$$
\bq = { \hbar \over m \a} {\rm Re} \ c_3, \quad
\bp = \hbar \left( {\rm Re} \ c_3 + {\rm Im} \ c_3 \right)
\eqno(2.17)
$$
Then the Gaussian may be written
$$
\eqalignno{
\exp \left( \ih c_1 x^2 + c_3 x \right)
&= \exp \left( - { m \a \over 2 \hbar} (1-i) (x -\bq)^2
+ \ih \bp x + { m \a \over 2 \hbar} (1-i) \bq^2 \right)
\cr
& \equiv \la x | \Psi_{\bp \bq} \ra 
\ e^{ { m \a \over 2 \hbar} (1-i) \bq^2 }
&(2.18) \cr}
$$
The propagator $K_{\bx}$ therefore has the form
$$
\eqalignno{
K_{\bx}(x_f,t|x_0,0) =  & N  \ \la x_f | \Psi_{\bp \bq} \ra 
\ e^{ { m \a \over 2 \hbar} (1-i) \bq^2 }
\cr
& \times \ \exp \left( \ih c_1 x_0^2 + c_4 x_0 + 
c_5 - a^2 \int_0^t ds \ \bx^2 (s) \right)
&(2.19) \cr }
$$
The generalized coherent states $ | \Psi_{\bp \bq} \ra $ depend on
$ \bx(t) $ only through $\bp$ and $\bq$, which are functionals
of $\bx (t) $. They are close to minimal uncertainty states,
satisfying, $\Delta p \Delta q = \hbar / \sqrt{2} $ [\cite{Dio3,HaZ}].

The desired form of the propagator is now obtained by inserting
(2.19) in (2.5), but reorganizing the functional integral over
$ \bx(t)$ into ordinary integrations over $\bp$ and $\bq$ and
functional integrations over remaining parts of $\bx(t)$.
This may be achieved by writing the functional integral
over $\bx (t)$ as
$$
\int {\cal D} \bx = \int dp dq \ \int {\cal D} \bx
\ \delta ( p - \bp) \ \delta (q -\bq)
\eqno(2.20)
$$
with $\bp$ and $\bq$ given in terms of $\bx$ by (2.17). We thus
obtain,
$$
\eqalignno{
J (x_f, y_f, t | x_0, y_0, 0 ) = & \int dp dq \ \int {\cal D} \bx
\ \delta ( p - \bp) \ \delta (q -\bq)
\ \la x_f | \Psi_{pq} \ra 
\ \la \Psi_{pq} | y_f \ra 
\ e^{ { m \a \over \hbar}  q^2 } 
\cr
& \times \ \exp 
\left( \ih c_1 x_0^2 - \ih c_1^* y_0^2 + c_4 x_0 + c_4^* y_0 
\right)
\cr
& \times \ \exp \left(
c_5 + c_5^*  - 2 a^2 \int_0^t ds \ \bx^2 (s) \right)
&(2.21) \cr}
$$
This may be written,
$$
J (x_f, y_f, t | x_0, y_0, 0 ) = \int dp dq 
\ f(p,q,t|x_0,y_0)
\ \la x_f | \Psi_{pq} \ra 
\ \la \Psi_{pq} | y_f \ra 
\eqno(2.22)
$$
where
$$
\eqalignno{
f(p,q,t|x_0,y_0) =
& \int {\cal D} \bx
\ \delta ( p - \bp) \ \delta (q -\bq)
\ e^{ { m \a \over \hbar}  q^2 } 
\cr
& \times \ \exp 
\left( \ih c_1 x_0^2 - \ih c_1^* y_0^2 + c_4 x_0 + c_4^* y_0 
\right)
\cr
& \times \exp \left(
c_5 + c_5^*  - 2 a^2 \int_0^t ds \ \bx^2 (s) \right)
&(2.23) \cr}
$$
We have clearly cast the result in the desired form. Folding an
arbitrary initial state into the expression for the density matrix
propagator (2.22), we obtain an expression of the desired form
(1.5), where $f(p,q,t)$ is given explicitly by,
$$
f(p,q,t) = \int dx_0 dy_0 \ f(p,q,t|x_0,y_0) \ \rho_0 (x_0,y_0)
\eqno(2.24)
$$
This is our first main result.

\head{\bf 3. The Phase Space Distribution Function}

It remains to evaluate the path integral expression (2.23).
To do this first notice that (2.23) may be written
$$
\eqalignno{
f(p,q,t|x_0,y_0) =&
\ \exp \left( \ih c_1 x_0^2 - \ih c_1^* y_0^2 + { m \a \over \hbar}  q^2 \right) 
\ \int dk dk' \ e^{\ih k p  + \ih k' q }
\cr
& \times 
\ \int {\cal D} \bx
\  \exp \left( - \ih k \bp - \ih k' \bq + c_4 x_0 + c_4^* y_0 \right)
\cr
& \quad \times \exp \left(
c_5 + c_5^*  - 2 a^2 \int_0^t ds \ \bx^2 (s) \right)
&(3.1) \cr}
$$
The functional integral over $\bx$ is a Gaussian, since $ c_5 $ is
quadratic in $\bx$ and $ \bp $, $\bq$ and $c_4$ are linear in $\bx$,
but it involves inverting the functional matrix contain in the
last exponential in (3.1), which does not look particularly easy.
However, we are saved from having to do this calculation by the
following observation. From Eq.(2.5) and Eq.(2.9) (for $\a t >> 1 $),
we see that
$$
\eqalignno{
J(x_f,y_f,t|x_0,y_0,0) =&
\exp \left( \ih c_1 (x_f^2 + x_0^2) - \ih c_1^* (y_f^2 +y_0^2)
\right)
\cr 
& \times \ \int {\cal D} \bx
\  \exp \left( c_3 x_f +c_3^* y_f + c_4 x_0 + c_4^* y_0 \right)
\cr
& \quad \times \exp \left(
c_5 + c_5^*  - 2 a^2 \int_0^t ds \ \bx^2 (s) \right)
&(3.2)\cr }
$$
This functional integral over $\bx$ in this expression
is very similar in form to (3.1) but
we already know what the answer is: it is Eq.(2.3).
In particular, equating (3.2) and (2.3), 
we obtain
$$
\eqalignno{
\int {\cal D} \bx
\  \exp \left( c_3 x_f +c_3^* y_f 
\right. & \left. 
+ c_4 x_0 + c_4^* y_0 
+ c_5 + c_5^*  - 2 a^2 \int_0^t ds \ \bx^2 (s) \right)
\cr
= & \exp \left(
{ i m \over 2 \hbar t} \left[ (x_f-x_0)^2 - (y_f-y_0)^2 \right]
\right.
\cr
& \quad \left.
- {a^2 t \over 6}  
\left[ (x_f -y_f)^2 + (x_f - y_f) (x_0 - y_0) + (x_0 - y_0)^2
\right] \right)
\cr
& \times \exp \left( - \ih c_1 (x_f^2 + x_0^2) + \ih c_1^* (y_f^2 +y_0^2)
\right)
&(3.3)\cr }
$$

Now the point is that the formula (3.3) is true for {\it arbitrary}
$x_f$, $y_f$. In particular, using (2.17), we see that
$$
c_3 x_f + c_3^* y_f = { m \a \over \hbar} \left[ x_f + y_f - i (x_f
-y_f) \right] \bq + \ih \bp (x_f - y_f)
\eqno(3.4)
$$
Hence the functional integral (3.3) is exactly the same as
the one appearing in (3.1) if, in (3.3), we make the substitutions
$$
(x_f - y_f) \ \ria \ -k , \quad
{ m \a \over \hbar} \left[ x_f + y_f - i (x_f -y_f) \right] 
\ \ria \ - \ih k'
\eqno(3.5)
$$
Inverting for $x_f$ and $y_f$, we therefore find that the functional
integral over $\bx (t) $ in (3.1) is equal to the right-hand 
side of (3.3) with 
$$
\eqalignno{
x_f &= - {(1+i) \over 2} k - { i  \over 2 m \a} k'
&(3.6) \cr
y_f &= { (1-i) \over 2} k - { i  \over 2 m \a} k'
&(3.7) \cr}
$$
Using this result, and
changing variables from $ k'$ to $ K = k + k' / m \a $ in (3.1),
we obtain
$$
\eqalignno{
f(p,q,t|x_0,y_0)=&
\exp \left( { m \a \over \hbar} q^2  
+ i { m X_0 \xi_0 \over \hbar t} - {a^2 t \over 6} \xi_0^2 \right)
\cr
& \times \int dk dK 
\ \exp \left( - \left( {a^2 t \over 6} - {m \a \over 4 \hbar} 
\right) k^2 - { m \a \over 4 \hbar} K^2 + \left( { m \a \over 2
\hbar} - { m \over 2 \hbar t} \right) k K \right)
\cr
& \quad \quad \times
\exp \left( \ih k \left( p - m \a q + {m X_0 \over
t} - i { \hbar a^2 t \over 6} \xi_0 \right) \right)
\cr 
& \quad \quad \times
\exp \left( \ih K \left( m \a q + i { m \xi_0 \over 2 t} \right)
\right)
&(3.8) \cr }
$$
where $X_0 = \half (x_0 + y_0) $, $\xi_0 = x_0 - y_0 $. This may
now be evaluated.

An alternative way of writing (3.8) is to carry out the same steps,
but to change variables in (3.1) from $k$, $k'$ to $x_f$, $y_f$,
with the formal result,
$$
\eqalignno{
f(p,q,t|x_0,y_0)= & \int dx_f dy_f
\exp \left( {m \a \over 2 \hbar} (1-i) (x_f -q)^2 - \ih px_f \right)
\cr & \times
\exp \left( {m \a \over 2 \hbar} (1+i) (y_f-q)^2 + \ih py_f \right)
J (x_f,y_f,t|x_0,y_0,0) 
&(3.9) \cr }
$$
Folding in the initial state via (2.24), we obtain,
$$
\eqalignno{
f(p,q,t)= & \int dx_f dy_f
\exp \left( {m \a \over 2 \hbar} (1-i) (x_f -q)^2 - \ih px_f \right)
\cr & \times
\exp \left( {m \a \over 2 \hbar} (1+i) (y_f-q)^2 + \ih py_f \right)
\rho_t (x_f, y_f)
&(3.10)\cr }
$$
which has the appearance of a formal inversion of the relation (1.5).

Because the coordinate transformation (3.6), (3.7) is complex
some attention to the integration contour is necessary. In
particular, $k$ and $k'$ are integrated along the real axis,
therefore $x_f +y_f$ is integrated along a purely imaginary contour
and $x_f -y_f$ along a real contour. More precisely, let
$ X= \half (x_f + y_f) $ and $ \xi = x_f - y_f $. Then (3.9) becomes
$$
\eqalignno{
f(p,q,t|x_0,y_0) = \int_{- i\infty}^{i \infty} dX \int_{- \infty}^{+\infty} 
d \xi 
& \ \exp\left( { m \a \over \hbar} \left( (X-q)^2 + {\xi^2 \over 4}
\right) - \ih m \a  \xi (X-q) - \ih p \xi \right)
\cr & \times
\ J (X + {\xi \over 2}, X - { \xi \over 2}, t | x_0, y_0, 0)
&(3.11) \cr }
$$
Explicitly, this integral reads,
$$
\eqalignno{
f(p,q,t|x_0,y_0) = & \int_{- i\infty}^{i \infty} dX \int_{- \infty}^{+\infty} 
d \xi 
\ \exp\left( { m \a \over \hbar} \left( (X-q)^2 + {\xi^2 \over 4}
\right) - \ih m \a  \xi (X-q) - \ih p \xi \right)
\cr & \times
\exp \left( { i m \over \hbar t} (X-X_0) (\xi - \xi_0)
- { 2 m \a^2 t \over 3 \hbar} ( \xi^2 + \xi \xi_0 + \xi_0^2 ) \right)
&(3.12) \cr }
$$
where $X_0$ and $\xi_0$ defined in the same way as $X$ and $\xi$.
The $X$ integral will clearly converge since the contour is along
the imaginary axis, and the $\xi$ integral will converge for
sufficiently large $\a t $.

Letting $X \ria X+ q $, the integral over $X$ is readily carried
out, with the result
$$
\eqalignno{
f(p,q,t|x_0,y_0) = \int d \xi
& \exp \left( - \ih p \xi + { i m \over \hbar t} (q-X_0) (\xi - \xi_0)
- { 2 m \a^2 t \over 3 \hbar} ( \xi^2 + \xi \xi_0 + \xi_0^2 ) \right)
\cr \times &
\exp \left( { m \a \over 4 \hbar} \left[ \xi^2
+ \left( \xi - { ( \xi - \xi_0) \over \a t } \right)^2 \right] \right)
&(3.13) \cr }
$$
The integral over $\xi $ may now be evaluated but it is not
necessary to do this, since the form of the answer is now clear.
For $\a t >> 1$, the terms in the second exponential are negligible
compared to the similiar terms in the first. Furthermore, the
remaining terms have the form of the Wigner transform of the
propagator [\cite{AnH,Wig}].
We thus have the simple result,
$$
f(p,q,t|x_0,y_0) \ \approx 
\ \int d \xi \ e^{ - \ih p \xi }
\ J( q,\xi, t | X_0, \xi_0, 0 )
\eqno(3.14) 
$$
Attaching an arbitrary initial density matrix, it then follows
from (2.24) that
$$
\eqalignno{
f(p,q,t) \ & \approx 
\ \int d \xi \ e^{ - \ih p \xi } \ \rho_t ( q+ \half \xi, q- \half \xi)
\cr &
= W_t (p,q)
&(3.15) \cr }
$$
That is, for $\a t >> 1$, $f(p,q,t)$ is the Wigner function of the
density operator at time $t$. This is the second main result of this paper.

From any of the above representations of $f(p,q,t)$ (other than (3.14)),
or from Ref.[\cite{HaZ}], it is straighforward to show
that $f(p,q,t)$ obeys the Fokker-Planck equation,
$$
{\partial f \over \partial t} = - {p \over m}
{ \partial f \over \partial q} + 
2 m \gamma k T { \partial^2 f \over \partial p^2} 
+ ( 2 \hbar \gamma k T)^{\half} { \partial^2 f \over \partial p
\partial q}
+ { \hbar \over 2 m } { \partial^2 f \over \partial q^2}
\eqno(3.16)
$$
As we have seen, $f(p,q,t)$ approaches the Wigner
function $W_t (p,q)$ for $\a t >> 1$, which obeys the
Fokker-Planck equation of classical Brownian motion:
$$
{\partial W \over \partial t} = - {p \over m}
{ \partial W \over \partial q} + 
2 m \gamma k T { \partial^2 W \over \partial p^2} 
\eqno(3.17)
$$
What happens is that the last two terms in Eq.(3.16) become
negligible for large $\a t $, as may be seen by studying
the Wigner function propagator (below).

\head{\bf 4. The Positivity of the Wigner Function}

We have shown that the density operator approaches the form
(1.5), where $f(p,q,t)$ is given by the Wigner function.
However, $f(p,q,t)$ is by construction positive, yet the Wigner
function is not guaranteed to be positive in general [\cite{Wig}].
What happens is that the Wigner function becomes strictly non-negative
after a period of time, under evolution according to (the Wigner
transform of) Eq.(1.1), as we now show.

The Wigner transform of the relation (2.1) yields,
$$
W_t (p,q) = \int dp_0 dq_0 \ K(p,q,t|p_0, q_0, 0 ) \ W_0 (p_0,q_0)
\eqno(4.1)
$$
where $K(p,q,t|p_0,q_0,0)$ is the Wigner function propagator, and is
given by [\cite{AnH}],
$$
K(p,q,t|p_0, q_0, 0 ) =
\exp \left( - \mu (p-p_0)^2 - \nu \left( q- q_0 - { p_0 t \over m} 
\right)^2 + \sigma (p-p_0) \left( q - q_0 - {p_0 t \over m} 
\right) \right)
\eqno(4.2)
$$
where, introducing $ D= 2 m \gamma k T $,
$$
\mu = {1 \over Dt}, \quad \nu = { 3 m^2 \over D t^3}, \quad
\sigma = { 3 m \over D t^2}
\eqno(4.3)
$$

It is well-known that the Wigner function may take negative values
only through oscillations in $\hbar$-sized regions of phase space,
and that it may be rendered positive by coarse-graining of such
a region. Considered for example, the smeared Wigner function
$$
W_H (p,q) = 2 \int dp' dq' \ \exp \left(
- { 2 \s_q^2 (p-p')^2 \over \hbar^2 } - { (q-q')^2 \over 2 \s_q^2 } \right)
\ W_{\rho} (p',q')
\eqno(4.4)
$$
This object is called the
Husimi function [\cite{Hus}]. It is equal to the expectation value of
the corresponding density operator in a coherent state
(of position width $\s_q$), $ \la p, q | \rho | p, q \ra $, 
so is non-negative.

Loosely speaking, what happens during time evolution
according to (4.1), is that, after a certain amount of time,
the propagator effectively smears the Wigner function over a region
of phase space greater than $\hbar$, and it becomes positive,
in the manner of (4.4). We will now show this explicitly.

Letting $p_0 \ria p_0 + p $ and $q_0 \ria q_0 + q - p_0 t / m $
in (4.1) yields,
$$
W_t (p,q) = \int dp_0 dq_0 \ \exp \left(
- \mu p_0^2 - \nu q_0^2 + \s p_0 q_0 \right)
\ W_0 (p_0 + p, q_0 + q - { p_0 t \over m} )
\eqno(4.5)
$$
The further transformation $ p_0 \ria p_0 + {\s \over 2 \mu} q_0 $
yields,
$$
W_t (p,q) = \int dp_0 dq_0 \ \exp \left(
- \mu p_0^2 - \b q_0^2  \right)
\ W_0 (p_0 + {\s \over 2 \mu} q_0 + p, q_0 + q - {p_0 t \over  m} )
\eqno(4.6)
$$
where $\b = \left( \nu - { \s^2 \over 4 \mu} \right) $. These two
transformations are canonical, and therefore the transformed Wigner
function appearing in the integrand of (4.6) is still the Wigner
function of some state (unitarily related to the original one).
Hence,
$$
W_t (p,q) = \int dp_0 dq_0 \ \exp \left(
- \mu p_0^2 - \b q_0^2  \right)
\ \tilde W_{pq} (p_0, q_0) 
\eqno(4.7)
$$
for some Wigner function ${\tilde W}_{pq}$ depending on $p,q$.
This may now be recast as the smearing of a Husimi function:
$$
\eqalignno{
W_t (p,q) = & \int dp' \ \exp \left( - { {p'}^2 \over
(\mu^{-1} - \hbar^2 \b) } \right)
\cr & \int dp_0 dq_0
\exp \left( - { (p'- p_0)^2 \over \hbar^2 \b} - \b q_0^2 \right)
\ \tilde W_{pq} (p_0,q_0)
&(4.8) \cr}
$$
The integral over $p_0$, $q_0$ is a Husimi function with
$\s_q^2 = 1 / (2 \b) $. Hence $W_t (p,q) \ge 0 $ provided 
the integral over $p'$ in (4.8) exists.
This will be the case if $\mu^{-1} > \hbar^2 \b $, that is, if
$$
t > \left( {\sqrt{3} \over 2} \right)^{\half}
\left( {\hbar \over \gamma k T} \right)^{\half}
\eqno(4.9)
$$
The Wigner function will therefore be non-negative for times greater
than the localization time (mulitplied by a number of order $1$).

\head{\bf 5. Discussion}

We have shown that
for times greater than the localization time, 
$( \hbar / \gamma k T)^{\half} $,
the density operator satisfying (1.1) approaches the form
$$
\rho = \int dp dq \ W_t (p,q) | \Psi_{pq} \ra \la \Psi_{pq} |
\eqno(5.1)
$$
where $W_t (p,q)$ is the Wigner function and the $ | \Psi_{pq} \ra $
are close to minimum uncertainty generalized coherent states.
The Wigner function is strictly non-negative for times greater
than the localization time (times a number of order $1$).

Di\'osi has also discussed the possibility of the phase space
diagonal form (1.5) under evolution according to the master equation
(1.1) [\cite{Dio8}]. His method was very different to ours, in that
he used the properties of the coherent states to regard (1.5)
as an expansion of the density operator. He found that such an
expansion is possible for times greater than the localization time,
times a number of order $1$, in tune with our results.

An advantage of deriving (5.1) using path integral methods,
rather than quantum state diffusion, is that it yields and explicit
expression for the phase space distribution function $f(p,q,t)$.
Another advantage is that it is
not obviously restricted to Markovian master equations. The quantum
state diffusion picture, in its current state of development, exists
only for systems described by a Markovian master equation. It may
exist in the non-Markovian case, but is yet to be developed. The
exact propagator for quantum Brownian motion, for quadratic
potentials, can be given in terms of a path integral [\cite{HPZ}], 
and is (mildly) non-Markovian. Since the method described here utitlizes
path integrals, rather than the quantum state diffusion picture,
there is a chance that our method may be valid in the non-Markovian
case also, but this is still to be investigated.

We have concentrated in this paper on the simplest possible model of
quantum Brownian motion: the free particle in a high temperature
environment with negligible dissipation $\gamma$. It is clear,
however, that remaining  in the context of a Markovian master
equation, it would be straighforward (although perhaps tedious) to
extend our considerations to the case of a harmonic oscillator with
non-trivial dissipation. In the quantum state diffusion picture
analysis, this case  was covered in Ref.[\cite{HaZ}] and we expect the path
integral treatment of the present paper to yield comparable results.

It is perhaps enlightening to comment on the various timescales
involved in a more general quantum Brownian motion model, and sketch
the expected general physical picture part of which is described by
the results of this paper.

In this paper, we have largely been concerned with the localization
time, $ ( \hbar / \gamma k T)^{\half} $, which is the timescale on
which an arbitrary initial density operator approaches the form
(1.5). The nomenclature ``localization time'' comes from the quantum
state diffusion picture, which was the picture first used to derive
some of the results described in this paper. It is so named because
it is the time scale on which an arbitrary initial wave function
becomes localized in phase space under evolution according to
Eq.(1.3) [\cite{HaZ,Dio3}].

Also relevant is the decoherence time, $ \hbar^2 / 
(\ell^2 m \gamma k T) $, which is the timescale on which the
off-diagonal terms of the density matrix are suppressed (in the
position representation) [\cite{Zur4}]. 
The decoherence time necessarily involves
a length scale $\ell$, which comes from the initial state. It could,
for example, be the separation of a superposition of localized
wave packets, and the decoherence time is then the time scale on
which the interference between these packets is suppressed.

If one is interested in emergent classicality for macroscopic
systems, it is appropriate to choose values order $1$ in c.g.s.
units, for $ \ell $, $T$, $m$ and $\gamma$.
The decoherence time is then typically much shorter than the localization
time. This is in turn typically much shorter than the relaxation time,
$\gamma^{-1}$, which is the time scale on which the system
approaches thermal equilibrium (when this is possible).

Hence the general picture we have is as follows. Suppose the initial
state of the system is a superposition of localized wave packets.
Then the interference terms between these wave packets is destroyed
on the decoherence timescale. After a few localization times the
density matrix subsequently approaches the phase space diagonal form
(1.5). After a much longer time of order the relaxation time, the
system reaches thermal equilibrium. Discussions of emergent
classicality usually concern times between the decoherence time and
the relaxation time, and it is this range of time which has been the
primary concern of this paper.

\head{\bf Acknowledgements}

We would like to thank
Todd Brun, 
Lajos Di\'osi, 
Juan Pablo Paz,
Ian Percival 
and Wojtek Zurek for useful conversations.

\references

\def\pr{{\sl Phys. Rev.\ }}
\def\prl{{\sl Phys. Rev. Lett.\ }}
\def\prep{{\sl Phys. Rep.\ }}
\def\jmp{{\sl J. Math. Phys.\ }}





\refis{AnH} C.Anastopoulos and J.J.Halliwell, \pr {\bf D51}, 6870 (1995).

\refis{AndH} A.Anderson and J.J.Halliwell, \pr {\bf 48}, 2753
(1993). 

\refis{Wig} N.Balazs and B.K.Jennings, \prep {\bf 104}, 347 (1984),
M.Hillery, R.F.O'Connell, M.O.Scully and E.P.Wigner, \prep {\bf
106}, 121 (1984); V.I.Tatarskii, {\sl Sov.Phys.Usp} {\bf 26}, 311 (1983).

\refis{BPSG} T.Brun, I.C.Percival, R.Schack and N.Gisin, 
QMW preprint (1996).

\refis{CaL} A.O.Caldeira and A.J.Leggett, {\sl Physica} {\bf 121A},
587 (1983).

\refis{CaM} C.M.Caves and G.J.Milburn, {\sl Phys.Rev.} {\bf A36}, 5543 (1987).


\refis{Dio2} L.Di\'osi, {\sl Phys.Lett.} {\bf 129A}, 419 (1988).

\refis{Dio3} L.Di\'osi, {\sl Phys.Lett} {\bf 132A}, 233 (1988).





\refis{Dio8} L.Di\'osi, {\sl Phys.Lett} {\bf A122}, 221 (1987).

\refis{DGHP} L.Di\'osi, N.Gisin, J.Halliwell and I.C.Percival,
{\sl Phys.Rev.Lett} {\bf 74}, 203 (1995). 


\refis{DoH} H.F.Dowker and J.J.Halliwell, {\sl Phys.Rev.} {\bf D46}, 1580
(1992)


\refis{Gal} M.R.Gallis, \pr {\bf A48}, 1023 (1993).

\refis{GaG} D.Gatarek and N.Gisin, {\sl J.Math.Phys.} {\bf 32},
2152 (1991).



\refis{GRW} G.C.Ghirardi, A.Rimini and T.Weber, {\sl Phys.Rev.} {\bf
D34}, 470 (1986); G.C.Ghirardi, P.Pearle amd A.Rimini, 
{\sl Phys.Rev.} {\bf A42}, 78 (1990).



\refis{GP1} N.Gisin and I.C. Percival, {\sl J.Phys.} {\bf A25},
5677 (1992); see also {\sl Phys. Lett.} {\bf A167}, 315 (1992).

\refis{GP2} N.Gisin and I.C.Percival, {\sl J.Phys.} {\bf A26},
2233 (1993).

\refis{GP3} N.Gisin and I.C.Percival,  {\sl J.Phys.} 
{\bf A26}, 2245 (1993).




\refis{HaZ} J.J.Halliwell and A.Zoupas, \pr {\bf D52}, 7294 (1995).

\refis{HaT} J.J.Halliwell and T.Yu, 
``Alternative Derivation of the Hu-Paz-Zhang Master
Equation of Quantum Brownian Motion'', accepted for publication in
{\sl Physical Review D}.

\refis{HPZ} B.L.Hu, J.P.Paz and Y.Zhang, \pr {\bf D45}, 2843
(1992); \pr {\bf D47}, 1576 (1993).

\refis{HZ} B.L.Hu and Y.Zhang, {\sl Mod.Phys.Lett.} {\bf A8},
3575 (1993).

\refis{Hus} K.Husimi, {\sl Proc.Phys.Math.Soc. Japan} {\bf 22},
264 (1940).


\refis{JoZ} E.Joos and H.D.Zeh, {\sl Z.Phys.} {\bf B59}, 223 (1985).



\refis{Men} M.B.Mensky, {\it Continous Measurement and Path
Integrals} (IOP Publishing, Bristol, 1993).


\refis{PHZ} J.P.Paz, S.Habib and W.Zurek, \pr {\bf D47}, 488 (1993).

\refis{PaZ} J.P.Paz and W.Zurek, \pr {\bf D48}, 2728 (1993).




\refis{QBM} G.S.Agarwal, \pr {\bf A3}, 828 (1971); \pr {\bf A4}, 739 (1971);
H.Dekker, \pr {\bf A16}, 2116 (1977); {\sl Phys.Rep.} {\bf 80}, 1 (1991);
G.W.Ford, M.Kac and P.Mazur, \jmp {\bf 6}, 504 (1965);
H.Grabert, P.Schramm, G-L. Ingold, \prep {\bf 168}, 115 (1988);
V.Hakim and V.Ambegaokar, \pr {\bf A32}, 423 (1985);
J.Schwinger, \jmp {\bf 2}, 407 (1961);
I.R.Senitzky, \pr {\bf 119}, 670 (1960).


\refis{SaG} Y.Salama and N.Gisin, {\sl Phys. Lett.} 
{\bf 181A}, 269, (1993).

\refis{Sch} L.Schulman, {\it Techniques and Applications of Path
Integration} (Wiley, New York, 1981).


\refis{UnZ} W.G.Unruh and W.Zurek, {\sl Phys.Rev.} {\bf D40}, 1071 (1989).


\refis{Zur} W.Zurek, {\sl Physics Today} {\bf 40}, 36 (1991)



\refis{Zur3b} W.Zurek, {\sl Phys.Rev.Lett.} {\bf 53}, 391 (1984).

\refis{Zur4} W.Zurek, in {\it Frontiers of Nonequilibrium Statistical
Physics}, edited by G.T.Moore and M.O.Scully (Plenum, 1986).

\refis{Zur5} W.Zurek, {\sl Prog.Theor.Phys.} {\bf 89}, 281 (1993);
and in, {\it Physical Origins of Time Asymmetry}, edited by 
J. J. Halliwell, J. Perez-Mercader and W. Zurek (Cambridge
University Press, Cambridge, 1994).

\refis{ZHP} W.Zurek, S.Habib and J.P.Paz, \prl {\bf 70},
1187 (1993).


\endreferences
\end